\documentclass[12pt]{article}
\usepackage{graphics}

\title{$CP$-Violation in the decay $B^0, \overline{B^0} \to \pi^+ \pi^- \gamma$}

\author{L. M. Sehgal and J. van Leusen \\ \it{Institute of Theoretical Physics, RWTH Aachen,} \\
\it{D-52056 Aachen, Germany}}

\date{}
\begin{document}

\maketitle

\begin{abstract}
The decay $\overline{B^0} \to \pi^+ \pi^- \gamma$ has a bremsstrahlung
component determined by the amplitude for $\overline{B^0} \to \pi^+ \pi^-$,
as well as a direct component determined by the penguin interaction
$V_{tb} V_{td}^* c_7 {\cal O}_7$. Interference of these amplitudes produces a
photon energy spectrum $\frac{d \Gamma}{d x} = \frac{a}{x} + b + c_1 x + c_2 x^2 + \cdots$
($x = \frac{2 E_{\gamma}}{m_B}$) where the terms $c_{1,2}$ contain a dependence on the phase
$\alpha_{eff} = \pi - {\rm arg}[(V_{tb}V_{td}^*)^*{\cal A}(\overline{B^0} \to \pi^+ \pi^-)]$.
We also examine the angular distribution of these decays, and show that in the
presence of strong phases, an untagged $B^0/\overline{B^0}$ beam can exhibit an
asymmetry between the $\pi^+$ and $\pi^-$ energy spectra.
\end{abstract}

\section{Introduction}

In this paper, we analyse the reaction $\overline{B^0} (B^0) \to \pi^+ \pi^- \gamma$,
with the aim of finding new ways of probing $CP$ violation in the non-leptonic
Hamiltonian, and in particular to test current assumptions about the weak and strong
phases in the amplitude for $\overline{B^0} \to \pi^+ \pi^-$. We will focus on
observables that can be measured in an untagged $\overline{B^0},B^0$ beam, which
are complementary to the time-dependent asymmetries in channels such as
$\overline{B^0}, B^0 \to \pi^+ \pi^-$ which are currently under study.

The amplitude for the decay of $\overline{B^0}$ into two charged pions and a
photon can be written as
\begin{equation}
{\cal A}(\overline{B^0} \rightarrow \pi^+ \pi^- \gamma) = {\cal A}_{brems} +
{\cal A}_{dir} \label{Ampwhole}
\end{equation}
where ${\cal A}_{brems}$ is the bremsstahlung amplitude and ${\cal A}_{dir}$ is the
direct emission amplitude. Our main interest will be in the continuum region of
$\pi^+ \pi^-$ invariant masses (large compared to the $\rho$-mass), and the possible
interference of the two terms in Eq. (\ref{Ampwhole}).

The bremsstrahlung amplitude is directly proportional to the amplitude for a
$\overline{B^0}$ decaying into two pions, the modulus of which is determined by the
measured branching ratio~\cite{BaBar:Belle}. Theoretically, the
$\overline{B^0} \rightarrow \pi^+ \pi^-$ decay amplitude can be written as~\cite{BBNS01}
\begin{equation}
{\cal A}(\overline{B^0} \rightarrow \pi^+ \pi^-) \sim V_{ub} V^*_{ud}T +
V_{cb}V^*_{cd}P \sim e^{-i \gamma} + \frac{P_{\pi \pi}}{T_{\pi \pi}}
\label{ABpp}
\end{equation}
where $\gamma = {\rm arg}(-V_{ud}V^*_{ub}/V_{cd}V^*_{cb})$, and $T$ and $P$ denote
the tree- and penguin-amplitudes, which can possess strong phases. (We follow the
notation of Ref.~\cite{BBNS01}, in which the phase of $\frac{P_{\pi \pi}}{T_{\pi \pi}}$
is estimated to be $\leq 10^{\circ}$). The $B^0 \to \pi^+ \pi^-$ amplitude is obtained by taking
the complex conjugate of the CKM factors in Eq. (\ref{ABpp}), leaving possible
strong phases unchanged. In the
experiments~\cite{BaBar:Belle}, one measures the time-dependent $CP$ asymmetry
\begin{equation}
A(t) = - S \sin (\Delta m_B t) + C \cos (\Delta m_B t)
\end{equation}
where
\begin{eqnarray}
S & = & \frac{2 {\rm Im} \lambda}{1+|\lambda |^2},
\,\, C = \frac{1-|\lambda |^2}{1+|\lambda |^2}, \\
\lambda & = & \frac{p}{q} \frac{{\cal A}(\overline{B^0} \rightarrow \pi^+ \pi^-)}{{\cal A}(B^0 \rightarrow \pi^+ \pi^-)}
= e^{-i 2 \beta} \frac{e^{-i \gamma}+ \frac{P_{\pi \pi}}{T_{\pi \pi}}}{e^{i \gamma}+ \frac{P_{\pi \pi}}{T_{\pi \pi}}}
\nonumber
\end{eqnarray}
In the limit of neglecting the penguin contribution, $P_{\pi \pi} \to 0$,
$\lambda = {\rm exp}[-i 2(\beta + \gamma)] = e^{2i \alpha}$, so that
$S = \sin 2 \alpha$, $C=0$. Theoretical considerations~\cite{BBNS01} suggest
$\left| \frac{P_{\pi \pi}}{T_{\pi \pi}} \right| \sim 0.28$. Present measurements
of $S$ and $C$ are yet inconclusive~\cite{BaBar:Belle, Nir}.

The direct emission amplitude ${\cal A}_{dir}$ is determined by the Hamiltonian
\begin{equation}
H_{peng} = V^*_{td} V_{tb} c_7 {\cal O}_7.
\end{equation}
This is the interaction which also leads to the exclusive decay
$B \to \rho \gamma$~\cite{StAlGr}. Here, however, we will be interested in the decay
$\overline{B^0} \to \pi^+ \pi^- \gamma$ for $\pi^+ \pi^-$ masses in the continuum
region, especially for large $s$ (or low photon energy). The Hamiltonian
$H_{peng}$ leads to
\begin{eqnarray}
{\cal A}_{dir} & = & \overline{E}_{dir}( \omega, \cos \theta) [ \epsilon \cdot p_+
k \cdot p_- - \epsilon \cdot p_- k \cdot p_+] \\ \nonumber
& & + i \overline{M}_{dir}( \omega, \cos \theta)
\epsilon_{\mu \nu \rho \sigma} \epsilon^{\mu} k^{\nu} p^{\rho}_+ p^{\sigma}_-
\end{eqnarray}
where the electric ($\overline{E}_{dir}$) and magnetic ($\overline{M}_{dir}$)
amplitudes depend on the two Dalitz plot coordinates: the photon energy
$\omega$ in the $\overline{B^0}$-meson rest frame and $\theta$, the angle
of the $\pi^+$ relative to the photon in the $\pi^+ \pi^-$ c.m. frame.

As long as the photon polarization is not observed, only the electric component of
the direct amplitude interferes with the bremsstrahlung amplitude. This interference
is in principle sensitive to the relative phase of
$T \lambda_u + P \lambda_c$ and $\lambda_t$ ($\lambda_i = V_{ib} V^*_{id}$) and
therefore could serve as a probe of the phase of ${\cal A}(\overline{B^0} \to \pi^+ \pi^-)$.

\section{Differential Decay Rate}

In accordance with the Low theorem the bremsstrahlung matrix element
is directly proportional to
the amplitude of $\overline{B^0} \to \pi^+ \pi^-$ on the mass shell:
\begin{equation}
{\cal A}_{brems} =
\frac{e {\cal A}(\overline{B^0} \to \pi^+ \pi^-)}{(k \cdot p_+)(k \cdot p_-)}
\left[ (\epsilon \cdot p_+)(k \cdot p_-) - (\epsilon \cdot p_-)(k \cdot p_+) \right]
\end{equation}

To obtain the direct amplitude ${\cal A}_{dir}$, we observe first that the operator
${\cal O}_7 \sim \overline{d} \sigma_{\mu \nu} (1+ \gamma_5) F^{\mu \nu} b$, and the
identity $\sigma_{\mu \nu} = \frac{i}{2} \epsilon_{\mu \nu \alpha \beta}
\sigma^{\alpha \beta} \gamma_5$ enables one to write
$\overline{E}_{dir}( \omega, \cos \theta) = \overline{M}_{dir}( \omega, \cos \theta)$.
One can write a multipole expansion for these direct amplitudes in the form~\cite{Lee:Wu}
\begin{eqnarray}
\overline{B^0}: \overline{E}_{dir}( \omega, \cos \theta) & = & E^{(1)}(\omega) + \cos \theta \frac{\beta \omega}{m_B} E^{(2)}(\omega)
+ \cdots \label{MultiExp} \\
B^0: E_{dir}( \omega, \cos \theta) & = & E^{(1)}(\omega) - \cos \theta \frac{\beta \omega}{m_B} E^{(2)}(\omega) + \cdots \nonumber
\end{eqnarray}
The simplest assumption is the dipole approximation
\begin{equation}
\overline{E}_{dir}( \omega, \cos \theta) = E^{(1)}(\omega)
\end{equation}
in which $\overline{E}_{dir}$ is independent of $\cos \theta$. In Section~\ref{SecAsy}
we will consider also consequences arising from a quadrupole term.

To get a dimensionless decay distribution we introduce
\begin{eqnarray}
x & = & \frac{2 \omega}{m_B} \\ 0 < & x & < 1 - \frac{4 m^2_{\pi}}{m^2_B} \nonumber
\end{eqnarray}
Then the differential branching ratio for the process
$\overline{B^0} \to \pi^+ \pi^- \gamma$ is:
\begin{eqnarray}
\frac{d Br}{dx d\cos \theta} & = & \frac{1}{512 \pi^3} \left( \frac{m_B}{2} \right)
\left( \frac{x}{2} \right)^3 \beta^3_{\pi} (1-x) \sin^2 \theta \label{EBrxt} \\
& & \times \left[ 2 |\overline{E}_{dir}|^2 + |\overline{E}_{brems}|^2 + 2 {\rm Re}
(\overline{E}^*_{dir} \overline{E}_{brems})\right] \nonumber
\end{eqnarray}
where
\begin{eqnarray}
|\overline{E}_{brems}| & = & e \sqrt{\frac{\pi}{m_B}}
\frac{64}{x^2 (1-\beta^2_{\pi} \cos^2 \theta)}
\sqrt{Br(\overline{B^0} \to \pi^+ \pi^-)} \label{EEbrms} \\
{\rm arg}(\overline{E}_{brems}) & = & - \gamma_{eff} = {\rm arg} \left( e^{-i \gamma}
+ \frac{P_{\pi \pi}}{T_{\pi \pi}} \right) \\
{\rm arg}(\overline{E}_{dir}) & = & {\rm arg}(V_{tb} V^*_{td}) = \beta\\
\beta^2_{\pi} & = & 1 - \frac{4 m^2_{\pi}}{s} \\
s & = & m_B^2(1-x)
\end{eqnarray}
where $Br(\overline{B^0} \to \pi^+ \pi^-) = 5.1 \times 10^{-6}$~\cite{BaBar:Belle},
$\beta = 24^{\circ}$~\cite{BaBar:Belle, Nir}
and  $P_{\pi \pi}/ T_{\pi \pi} = 0.28$, with a negligible strong phase~\cite{BBNS01}.

The bremsstrahlung part of the decay distribution populates preferentially the
region of small $x$ and $|\cos \theta | \sim 1$. Far from this region, the spectrum
is determined by the direct term $| \overline{E}_{dir}|^2$. In principle, a fit
to the Dalitz plot can determine the scale and the shape of the direct amplitude.

\begin{figure}
\center
\makebox[6.5cm]{
\resizebox{6.5cm}{4.9cm}
{\includegraphics*{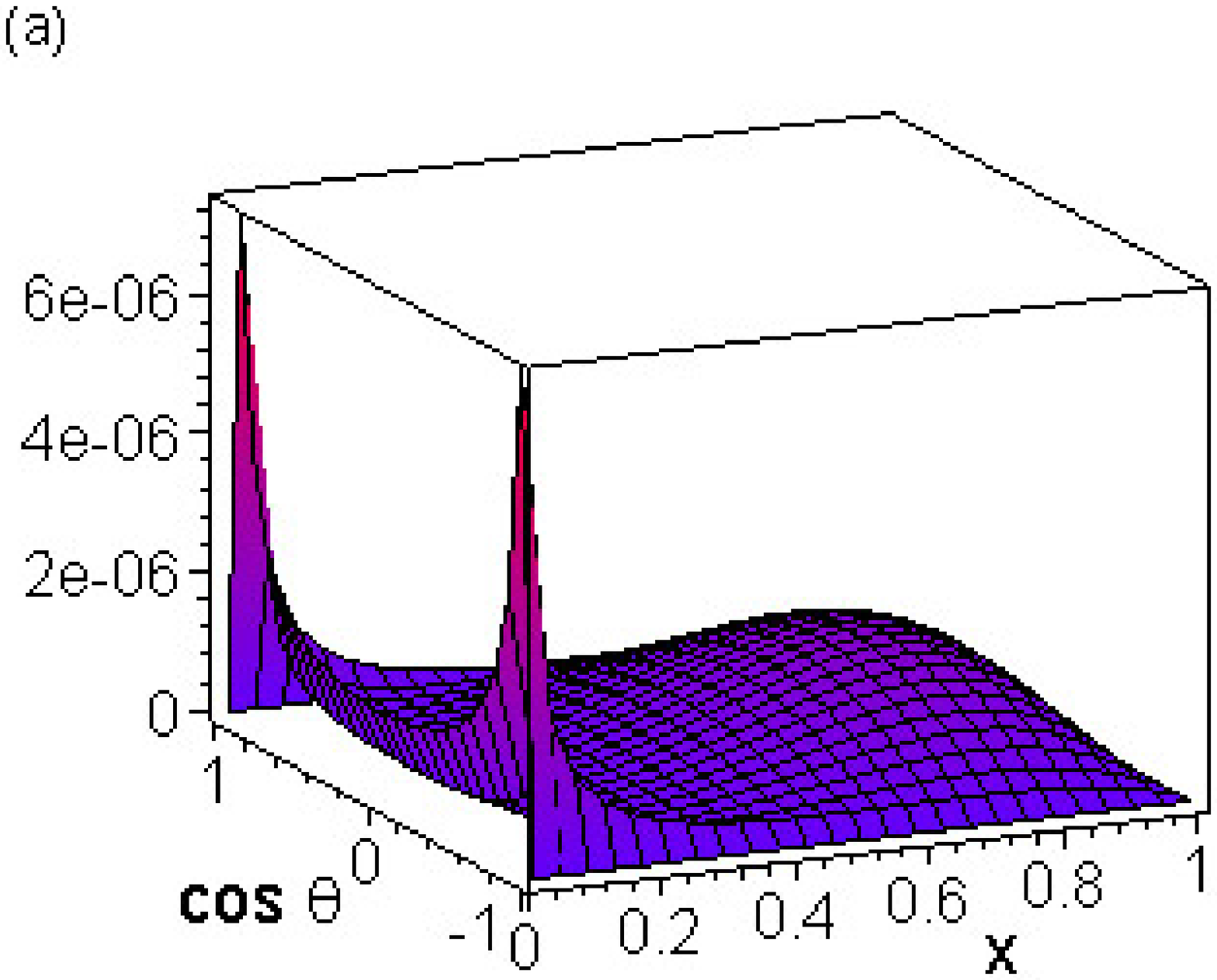}}
}
\makebox[6.5cm]{
\resizebox{6.5cm}{4.9cm}
{\includegraphics*{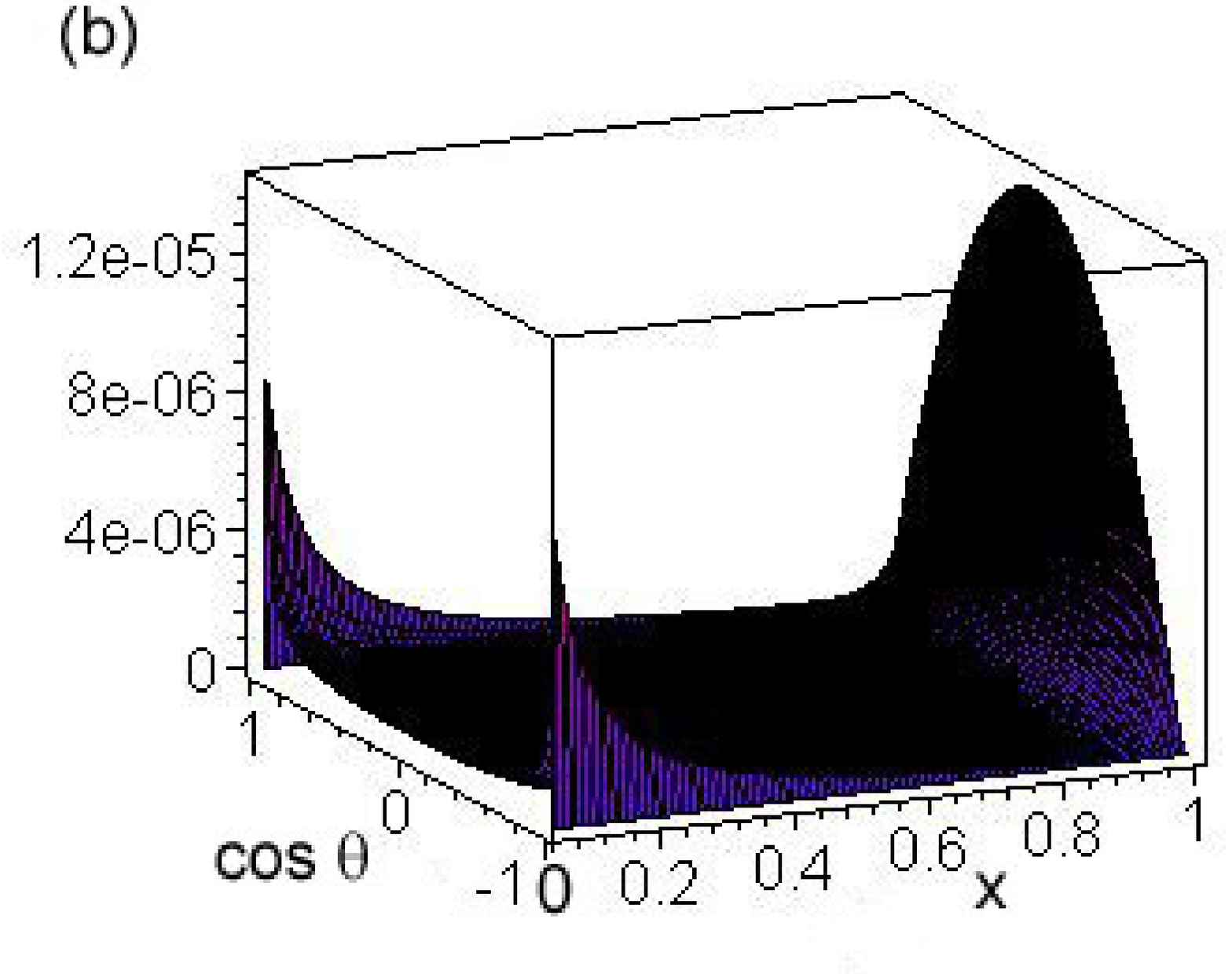}}
}
\caption{The differential branching ratio $\frac{d Br}{dx d \cos \theta}$ for a) $E^{(1)} = const.$ and
b) $E^{(1)} \sim 1/s$. Parameters in both plots: $\alpha = 80^{\circ}$ and
$Br_{dir} = 10^{-6}$.  \label{BRxandtheta}}
\end{figure}

Fig.~\ref{BRxandtheta} shows the two-dimensional decay distribution, for an assumed
direct branching ratio $Br_{dir} = 10^{-6}$ and two choices of form factor
$E^{(1)} = const.$ and $E^{(1)} \sim 1/s$. In the dipole
approximation the distribution is symmetric with respect to $\cos \theta$ and
identical for $\overline{B^0}$ and $B^0$ decay.

\section{Photon Energy Spectrum}

Integrating over the variable $\cos \theta$, the branching ratio, for small
energies $x$, can be written in the form:
\begin{equation}
\frac{d Br}{dx} = \frac{a}{x} + b + c_1 x + c_2 x^2 + \cdots \label{EPhoto}
\end{equation}
In agreement with the Low theorem, the coefficients $a$ and $b$ are determined entirely
by bremsstrahlung whereas $c_1$, $c_2$ depend on the interference of bremsstrahlung
and direct emission and therefore contain information about the
relative phase of $T \lambda_u + P \lambda_c$ and $\lambda_t$ given by
$\beta+\gamma_{eff} \equiv \pi - \alpha_{eff}$. Higher order terms in $x$
($c_i$, $i=3, 4, \dots$) contain pure direct emission, in addition to
interference terms involving higher multipoles. Numerical estimates of the
expansion parameters are given in table~\ref{tableparam}.

\begin{figure}
\center
\makebox[6.5cm]{
\resizebox{6.5cm}{4.9cm}
{\includegraphics*{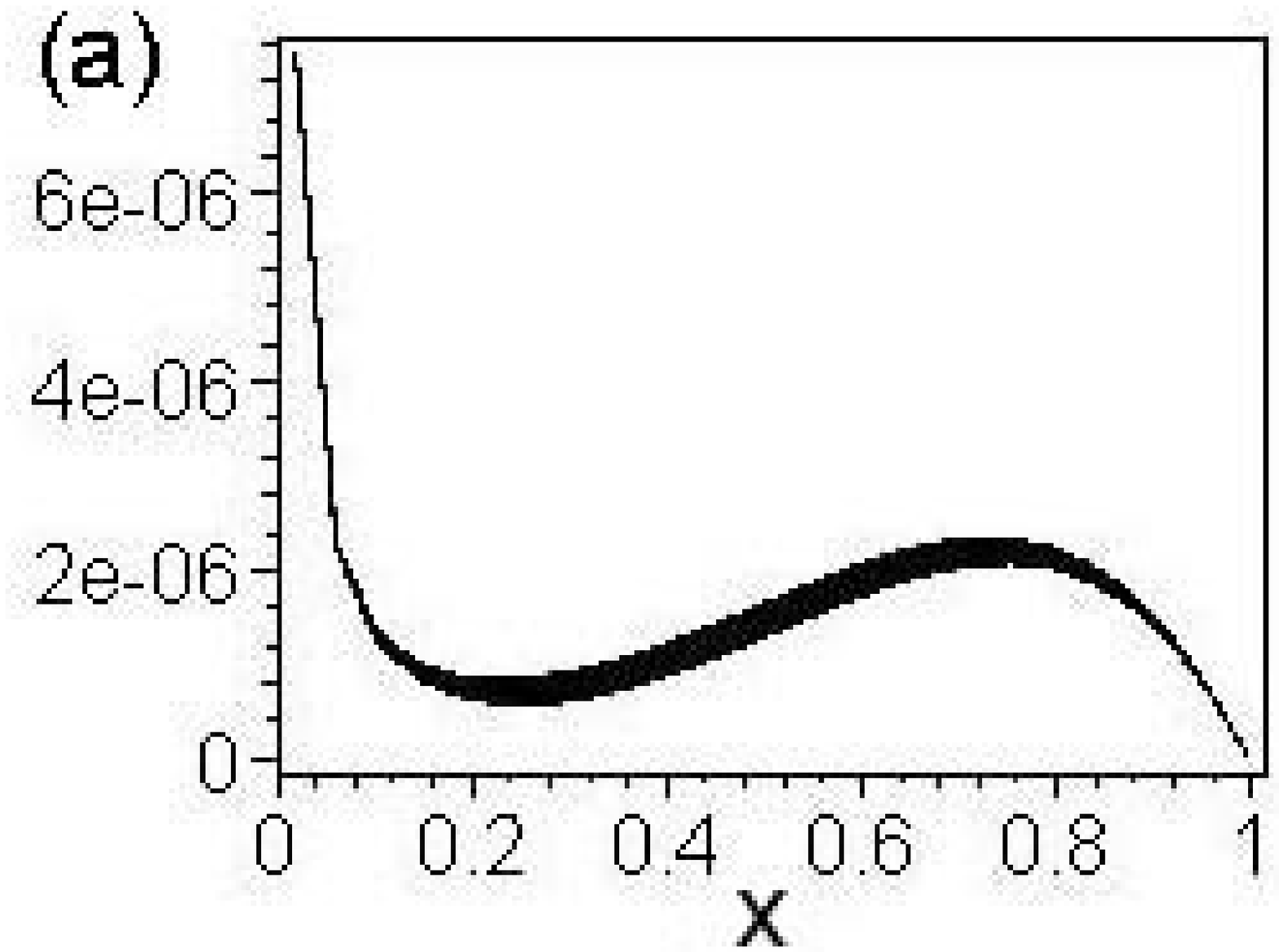}}
}
\makebox[6.5cm]{
\resizebox{6.5cm}{4.9cm}
{\includegraphics*{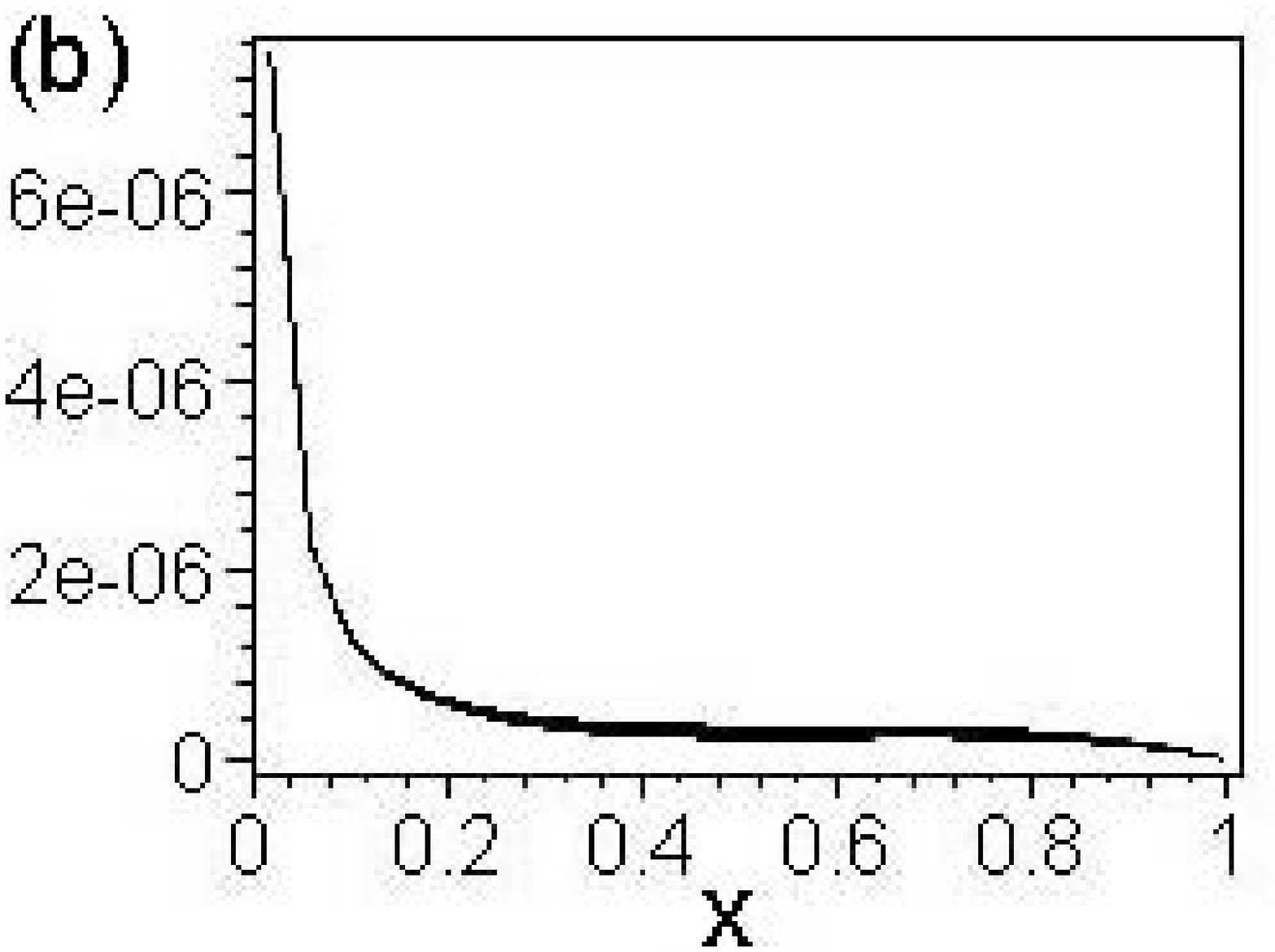}}
}
\makebox[6.5cm]{
\resizebox{6.5cm}{4.9cm}
{\includegraphics*{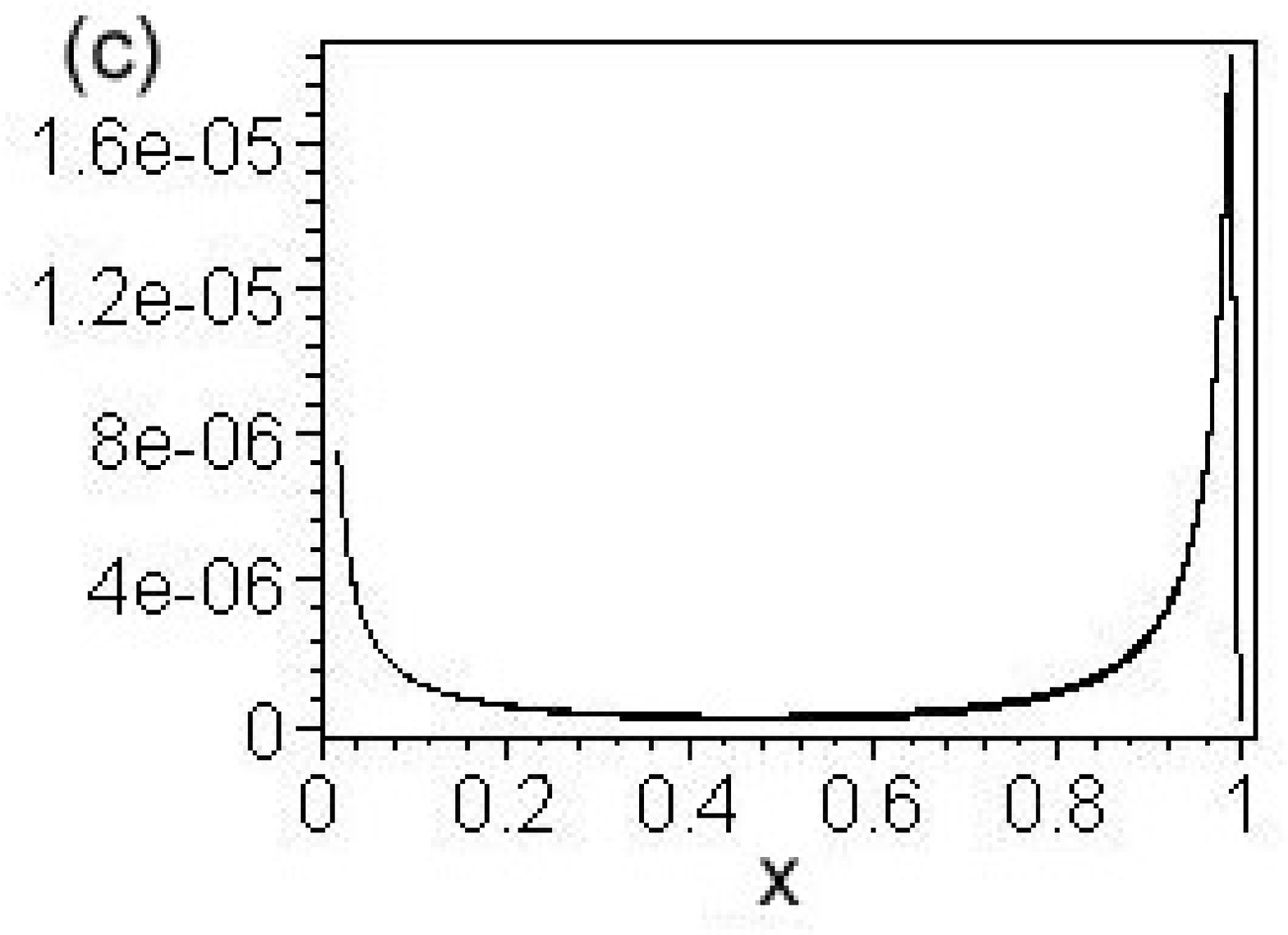}}
}
\makebox[6.5cm]{
\resizebox{6.5cm}{4.9cm}
{\includegraphics*{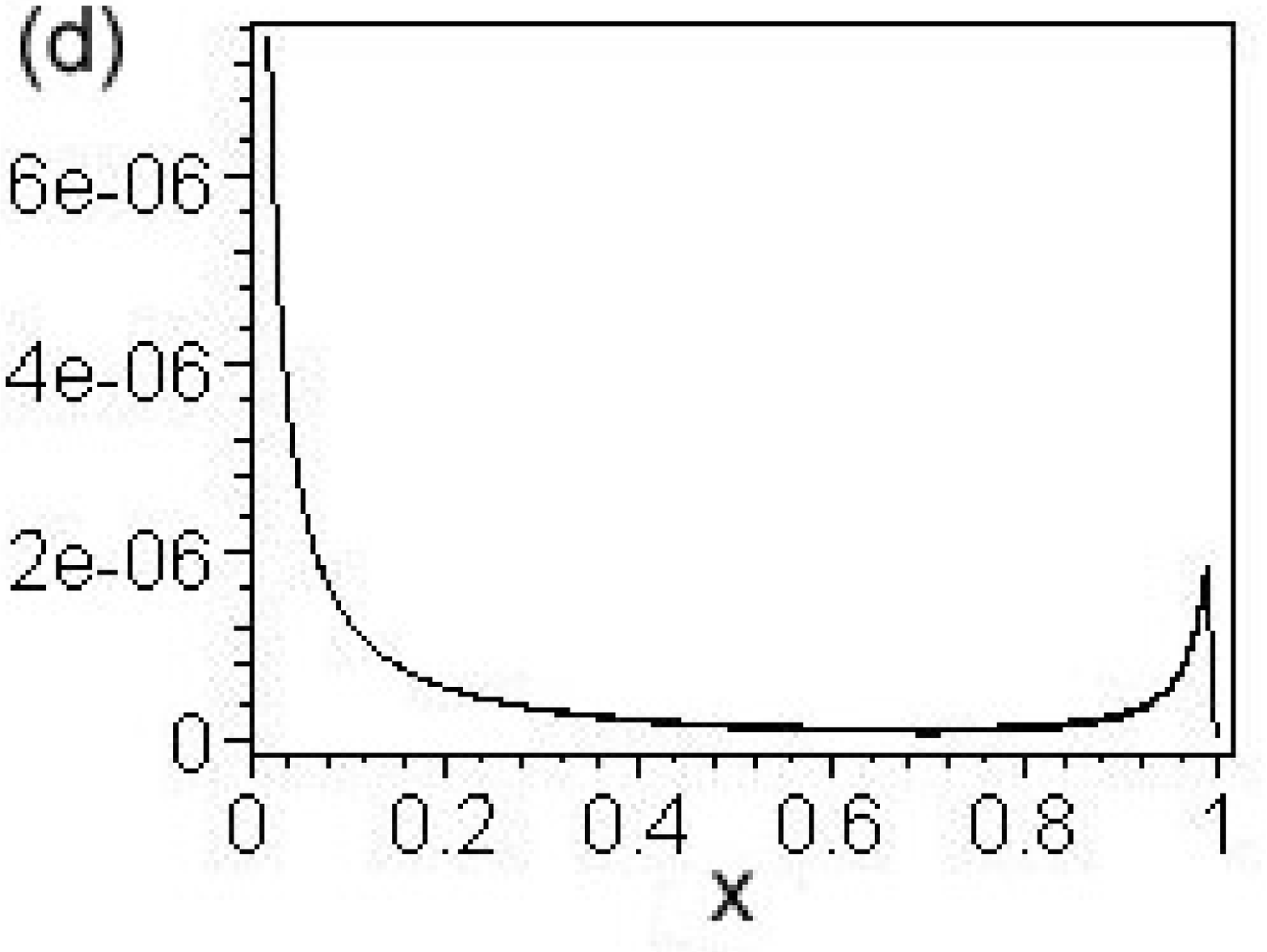}}
}
\caption{$\frac{dBr}{dx}$ for $E^{(1)} = const.$: a) $Br_{dir} = 10^{-6}$,
b) $Br_{dir} = 10^{-7}$ and $E^{(1)} \sim 1/s$: c) $Br_{dir} = 10^{-6}$,
d) $Br_{dir} = 10^{-7}$. The thickness of the lines represents the variation of
$\alpha$ from $60^{\circ}$ to $120^{\circ}$. \label{BRx}}
\end{figure}

In figure~\ref{BRx} we show typical photon energy spectra, for direct branching
ratios $10^{-6}$ or $10^{-7}$, and two choices of the form factor in $\overline{E}_{dir}$.
The phase $\alpha$ is allowed to vary between $60^{\circ}$ and $120^{\circ}$. As
expected, the sensitivity to $\alpha$ depends on the degree of overlap between the
bremsstrahlung and direct amplitudes. The expansion in Eq. (\ref{EPhoto}) turns out
to be a good description in the region $x \leq 0.25$. (Note that the resonant
contribution $\overline{B^0} \to \rho \gamma \to \pi^+ \pi^- \gamma$ would appear
as a spike in $\frac{d Br}{dx}$ at $x = 1-\frac{m^2_{\rho}}{m_B^2}$.)

In the absence of strong phases, the photon energy spectrum (\ref{EPhoto}) holds
for $B^0$ as well as $\overline{B^0}$ decay, and the results in table~\ref{tableparam}
and figure~\ref{BRx} are, in that limit, valid for an untagged $B^0$, $\overline{B^0}$
beam as well.

\section{Asymmetry in the Angular Distribution \label{SecAsy}}

\begin{figure}
\center
\makebox[6.5cm]{
\resizebox{6.5cm}{4.9cm}
{\includegraphics*{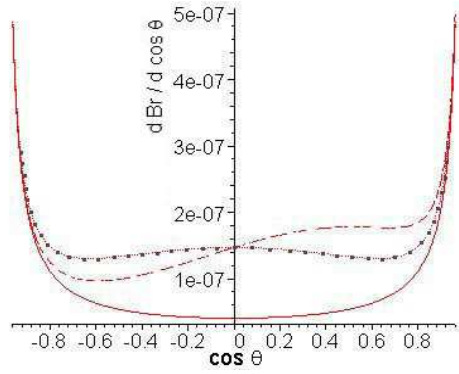}}
}
\caption{$\frac{d Br}{d \cos \theta}$ for $\overline{B^0} \to \pi^+ \pi^- \gamma$,
assuming $E^{(i)} = const.$ and a lower limit $\omega_{min} = 50 MeV$ for photon
energy. The solid line represents pure bremsstrahlung, the dotted line an additional
dipole contribution and the dashed line combines bremsstrahlung, dipole and quadrupole
contributions.  \label{CompDiQuad}}
\end{figure}

In the dipole approximation, the angular distribution is the same for $B^0$ and
$\overline{B^0}$ decay and is symmetric in $\cos \theta$ (c.f. Eq. (\ref{EBrxt})).
In the presence of an additional quadrupole term in the direct emission amplitude the
multipole expansion for $\overline{B^0}$ and $B^0$ differs by a sign, as shown in
Eq. (\ref{MultiExp}). For a numerical estimate of $E^{(2)}/E^{(1)}$, we
take~\cite{Belle}
\begin{equation}
\frac{Br(B \to \pi^+ \pi^- \gamma; E^{(2)})}
{Br(B \to \pi^+ \pi^- \gamma; E^{(1)})} \approx
\frac{Br(B \to K_2 \gamma)}{Br(B \to K^* \gamma)} \approx 0.25
\end{equation}
The result is plotted in fig.~\ref{CompDiQuad}, which is obtained from Eq. (\ref{EBrxt})
after integrating over photon energies $\omega > 50 MeV$. While the distribution
$\frac{d Br}{d \cos \theta}$ is symmetric in a dipole approximation, it develops
a forward-backward asymmetry in the presence of a quadrupole term.

Comparing $\overline{B^0}$ to $B^0$ we see that as long as strong phases are absent,
\begin{equation}
\frac{d Br(B^0 \to \pi^+ \pi^- \gamma)}{d \cos \theta} =
\frac{d Br(\overline{B^0} \to \pi^+ \pi^- \gamma)}{d \cos \theta}
(\cos \theta \to - \cos \theta),
\end{equation}
so that the decay of an untagged beam would be symmetric in $\cos \theta$.

This conclusion changes, however, if strong phases are not negligible.
Let us associate strong phases $\delta_0$, $\delta_1(s)$ and $\delta_2(s)$ with the
bremsstrahlung, direct dipole, and direct quadrupole terms. Notice that the bremsstrahlung
phase $\delta_0$ is the phase of the $\overline{B^0} \to \pi^+ \pi^-$ amplitude, and
therefore describes a $2 \pi$ state with $L  = 0$ and invariant mass $m_B$. By contrast, the
phases $\delta_1(s)$ and $\delta_2(s)$ describe $L=1$ and $L=2$ states of a $2 \pi$ system
with invariant mass $s$.
As a net result, the relative phase of the bremsstrahlung and direct amplitudes in Eq.
(\ref{EBrxt}) may be written as $\pi - \alpha_{eff}+\delta_{str}$ in $\overline{B^0}$
decay, and $-\pi + \alpha_{eff}+\delta_{str}$ in $B^0$ decay, where $\delta_{str}$ denotes
some effective combination of $\delta_0$, $\delta_1$ and $\delta_2$.

\begin{figure}
\center
\makebox[6.5cm]{
\resizebox{6.5cm}{4.9cm}
{\includegraphics*{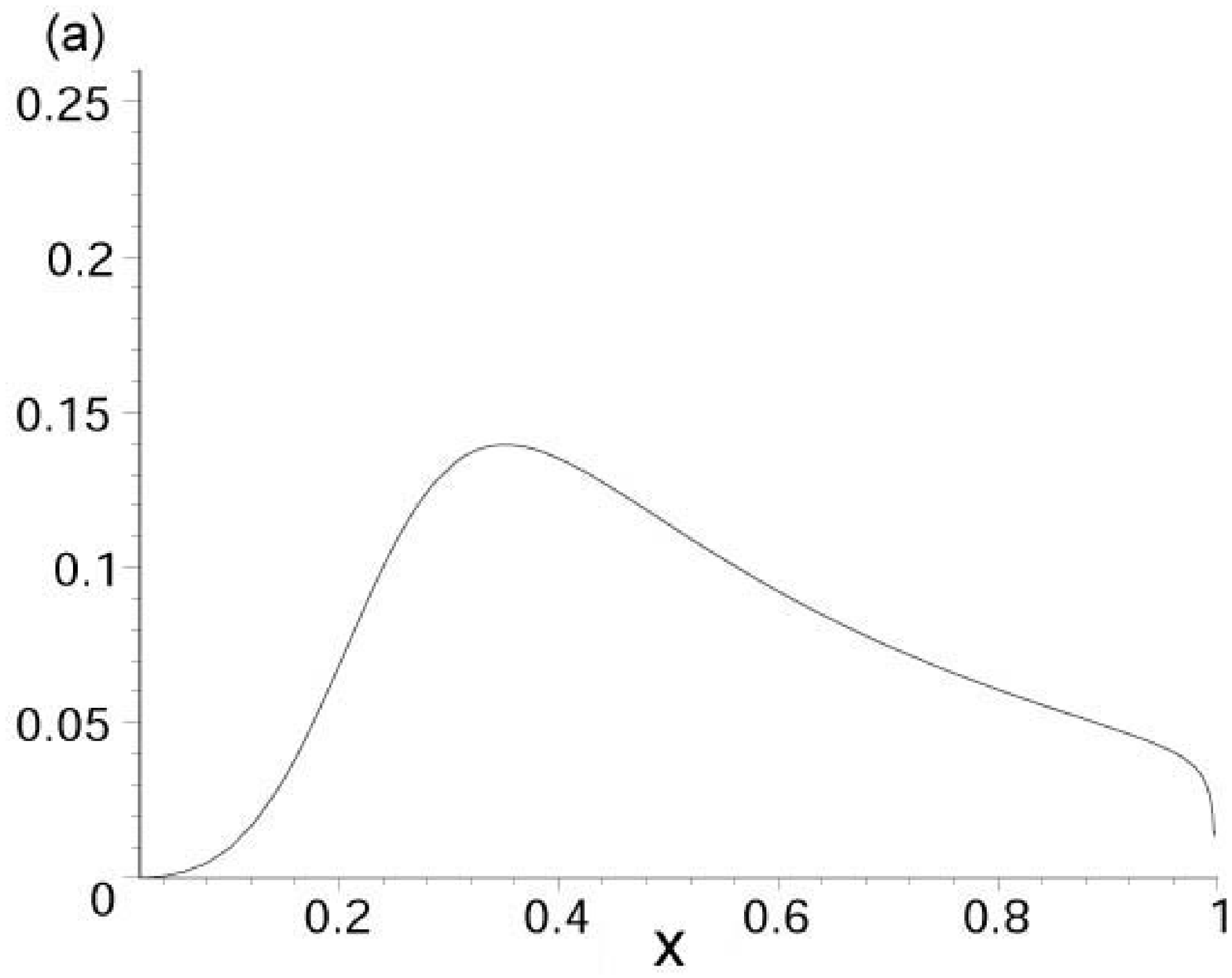}}
}
\makebox[6.5cm]{
\resizebox{6.5cm}{4.9cm}
{\includegraphics*{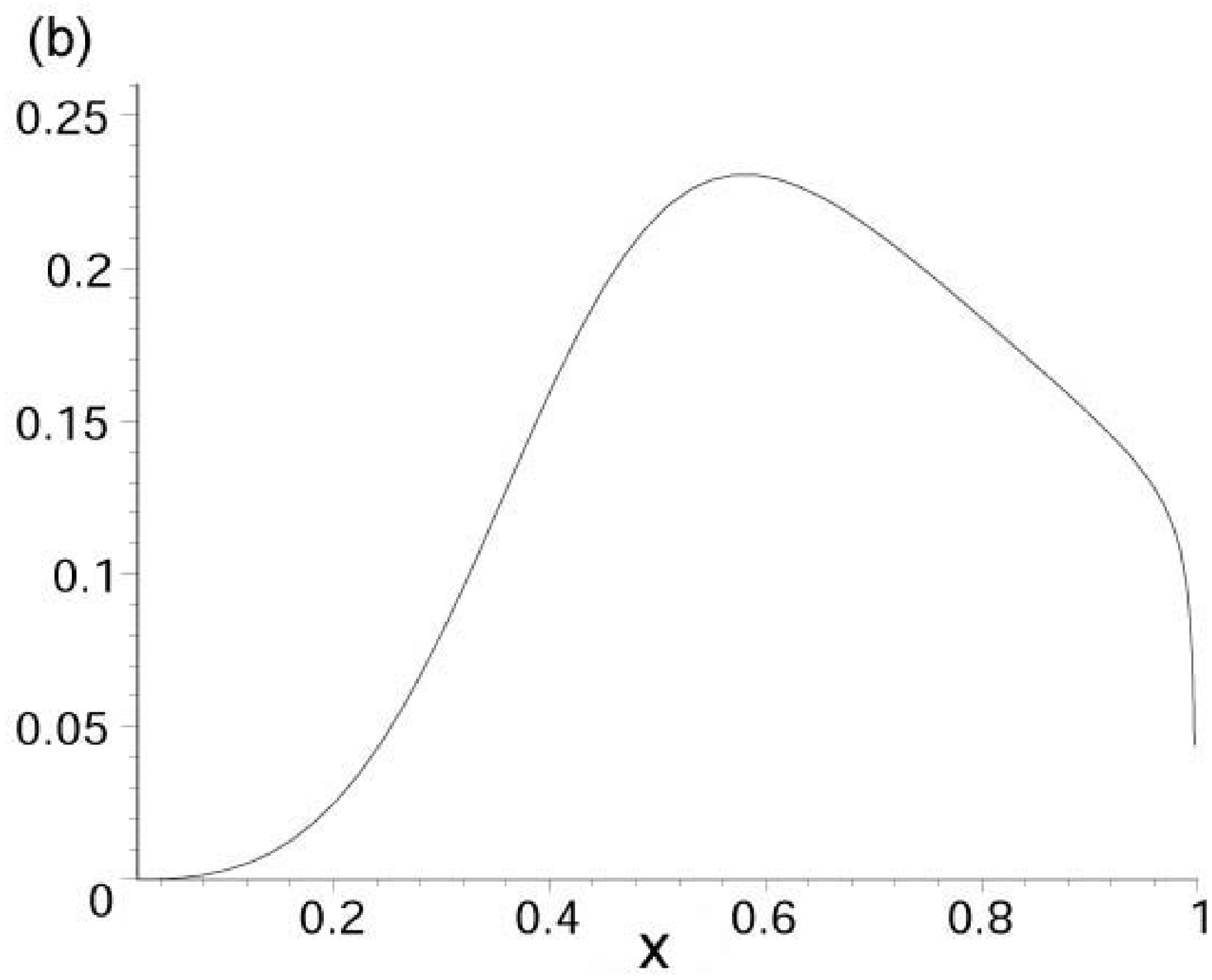}}
}
\makebox[6.5cm]{
\resizebox{6.5cm}{4.9cm}
{\includegraphics*{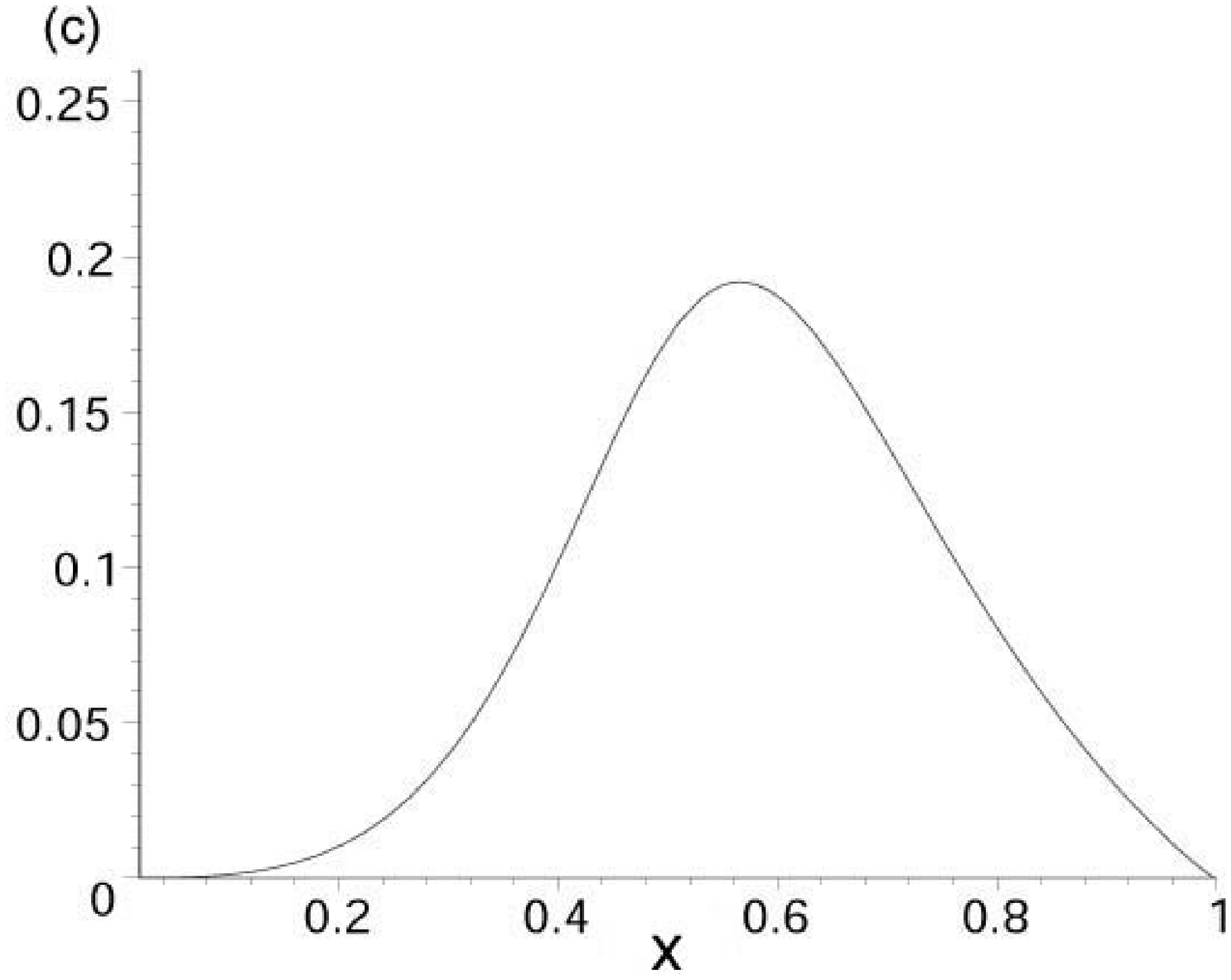}}
}
\makebox[6.5cm]{
\resizebox{6.5cm}{4.9cm}
{\includegraphics*{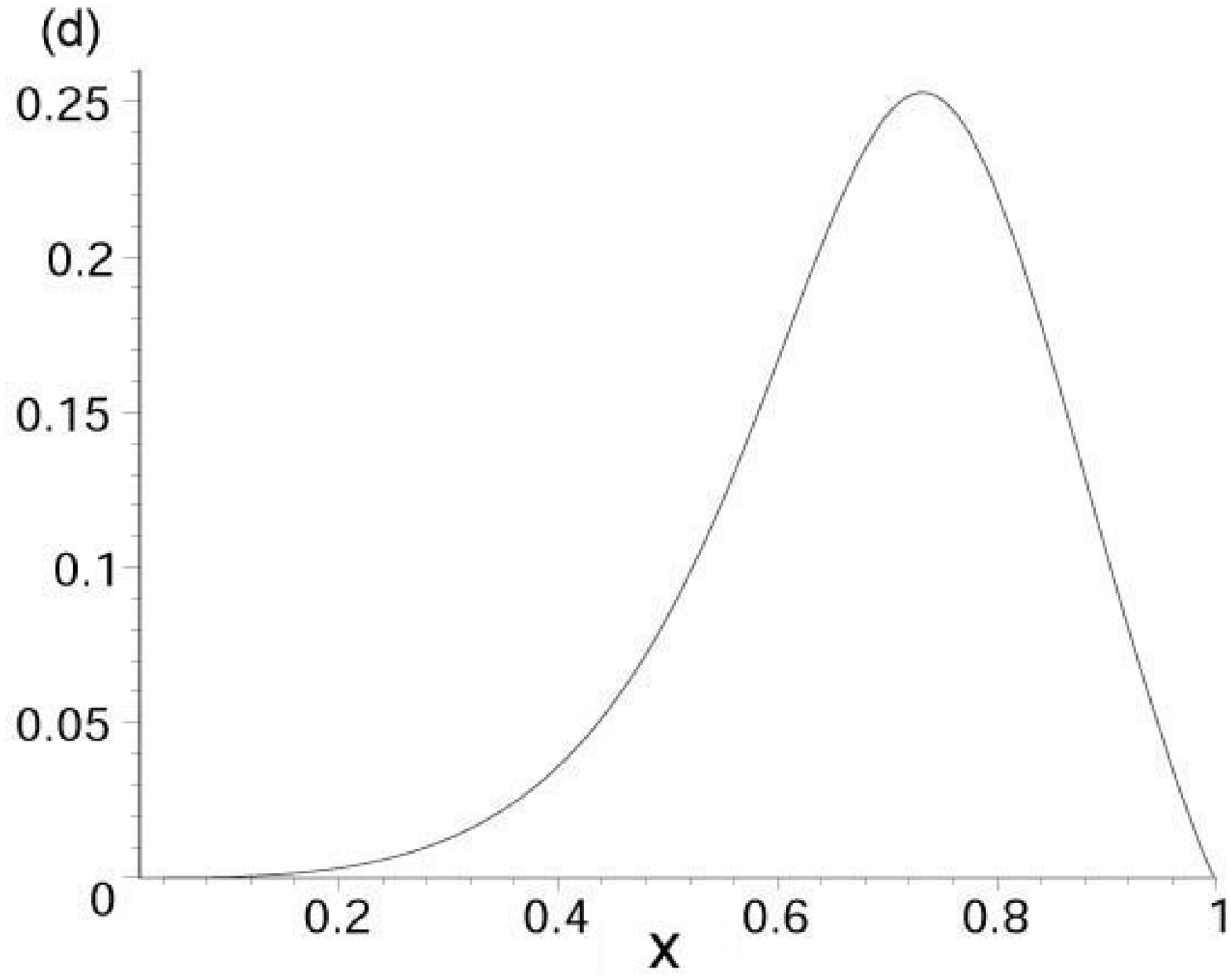}}
}
\caption{The function $f(x)$ (Eq. (\ref{EAsy})) describing the forward-backward
asymmetry in an untagged $B^0/ \overline{B^0}$ beam.
a) $Br_{dir} = 10^{-6}$, $E^{(i)} = const.$,
b) $Br_{dir} = 10^{-7}$, $E^{(i)} = const.$,
c) $Br_{dir} = 10^{-6}$, $E^{(i)} \sim 1/s$
d) $Br_{dir} = 10^{-7}$, $E^{(i)} \sim 1/s$.\label{AsyPlot}}
\end{figure}

Defining
\begin{eqnarray}
{\cal F}(x) & = & \int^1_0 \frac{d Br}{dx d \cos \theta} d \cos \theta \\
{\cal B}(x) & = & \int^0_{-1} \frac{d Br}{dx d \cos \theta} d \cos \theta \nonumber
\end{eqnarray}
we obtain the forward-backward asymmetry in a mixture of $B^0$ and $\overline{B^0}$:
\begin{equation}
Asy(x) = \frac{{\cal F - B}}{{\cal F + B}} = f(x) \cdot \sin \delta_{str}
\cdot \sin \alpha_{eff} \label{EAsy}
\end{equation}
Note, that the forward-backward asymmetry in (untagged) $\overline{B^0}, B^0 \to \pi^+ \pi^- \gamma$
decay is equivalent to an asymmetry in the energy spectrum of $\pi^+$ and $\pi^-$
in the $B$ meson rest frame. The
function $f(x)$ is plotted in figure~\ref{AsyPlot}.
Thus $Asy(x)$ is a signal of $CP$-violation, that is present even in an untagged
$\overline{B^0}/B^0$ mixture, and requires $\alpha_{eff} \neq 0$ and
$\delta_{str} \neq 0$, in addition to a quadrupole term in the direct electric
amplitude.

\section{Summary}

We have studied observables in the decay $B^0, \overline{B^0} \to \pi^+ \pi^- \gamma$
that do not require tagging or measurement of time-dependence, but which nevertheless
probe weak and strong phases appearing in the decay amplitude. Interference of the
bremsstrahlung and direct components affects the linear and quadratic terms in the
photon energy spectrum Eq. (\ref{EPhoto}), with potential sensitivity to the phase
${\rm arg}[(V_{tb}V_{td}^*)^*{\cal A}(\overline{B^0} \to \pi^+ \pi^-)]$. In the
presence of non-trivial strong phases, there is a difference in the $\pi^+$ and
$\pi^-$ energy spectra even for an untagged $B^0/ \overline{B^0}$ beam, or, equivalently,
a forward-backward asymmetry of the $\pi^+$ relative to the photon direction.

\begin{table}
\center
\begin{tabular}{|c|c|}
\hline
& $E^{(1)} = const.$ \\
\hline
$a$ & $1.45 \times 10^{-7}$ \\
$b$ & $-1.69 \times 10^{-7}$ \\
$c_1$ & $1.16 \times 10^{-8} + 1.67 \times 10^{-6} \sqrt{\frac{Br_{dir}}{10^{-6}}} \cos \alpha_{eff}$\\
$c_2$ & $3.88 \times 10^{-9} - 1.69 \times 10^{-6} \sqrt{\frac{Br_{dir}}{10^{-6}}} \cos \alpha_{eff}$\\
\hline
& $E^{(1)} \sim 1/s$ \\
\hline
$a$ & $1.45 \times 10^{-7}$ \\
$b$ & $-1.69 \times 10^{-7}$ \\
$c_1$ & $1.16 \times 10^{-8} + 2.19 \times 10^{-7} \sqrt{\frac{Br_{dir}}{10^{-6}}} \cos \alpha_{eff}$ \\
$c_2$ & $3.88 \times 10^{-9} - 2.27 \times 10^{-9} \sqrt{\frac{Br_{dir}}{10^{-6}}} \cos \alpha_{eff}$\\
\hline
\end{tabular}
\caption{The expansion parameters $a$, $b$, $c_1$, $c_2$ in the photon energy spectrum
(Eq. (\ref{EPhoto})) for two form factor models. The parameters $a$ and $b$ depend only on
bremsstrahlung and are therefore the same in both models. \label{tableparam}}
\end{table}

\end{document}